\documentclass[a4paper]{jpconf}
\usepackage{graphicx}
\begin{document}
\title{ A united model  for the cosmic ray energy spectra and anisotropy in the energy range 100 --100 000 GeV
}

\author{V I  Zatsepin, A D  Panov, N V Sokolskaya}

\address{Skobeltsyn Institute of Nuclear Physics, Moscow State University, Russia}

\ead{viza@dec1.sinp.msu.ru}
\begin{abstract}
We propose a model  where a supernova  explodes in some vicinity of our solar system (some tens of parsecs)
in the recent past (some tens of thousands years) with the energy release in cosmic rays  of order of $ 10 ^ {51} $ erg.
The flux from this supernova is added to an isotropic flux from other sources. 
We consider the case where the Sun's location  is not  in some typical for Our Galaxy average environment, but in the Local Superbubble about 100 pc across,
 in which the diffusion coefficient $ D (E) = D_0 \times E ^ {0.6} $ , with the value of $ D_0 \sim 10 ^ {25} $ cm $ ^ 2 $ s $ ^ {-1} $.
We describe the energy dependence of the anisotropy of cosmic rays in the TeV region,  together with the observed features of the energy spectrum of protons 
found in direct measurements. 
Our model provides a natural explanation to the hardening of the proton spectrum at 200 GeV, together with the observed steepening of the spectrum above 50 TeV.
\end{abstract}
\section{Introduction}
Notwithstanding the numerous attempts undertaken in recent years there  is not a satisfactory explanation of the energy dependence of cosmic ray  anisotropy
 amplitude in the frame 
of  existing paradigm \cite{blasi}, \cite{sveshnikova}. 
When addressing the problem the following is normally assumed:
a) The experimental data reflect a stationary picture;
b) The  CR propagation is determined by the diffusion of charged particles in the magnetic fields of our Galaxy;
c) The diffusion coefficient depends on the particles rigidity as:  $D (R) = D_0 \times R ^\delta $, where $ \delta $ is a parameter determining the
properties of the medium in the region where CR
 propagate in the Galaxy. It is assumed that $ D_0 $ and $\delta $ are the same
  throughout the Galaxy;
d) The distribution of supernovae as a dominant source of cosmic rays follows the spatial distribution of a galactic matter density for the remote sources at
distances greater than 1--1.5 kpc. To specify the details the Galaxy's spiral arms are taken into account. The observed distribution of the nearby sources (in space and time)
are also considered. 

In this paper we propose a model where the observed anisotropy is related to a non-stationary process in a nearby vicinity of the Sun.
There is some evidence that the Sun's environment  is not representative for typical on  average galactic conditions, and the Sun is in a  bubble of
 hot gas with a low density. To create this bubble the energy of many supernova explosions
in recent tens of millions years is needed \cite {streitmatter}. It is a reasonable suggestion that the CR energy density  is  significantly
higher in the bubble than in the the Galaxy on average. Even now there are stormy events of a neaby stars formation
 \cite{breitschwerdt}. We shall consider that the cosmic ray flux from a local source 
 is added to an isotropic background from a great number of other supernovae exploded in the Galactic disk (see \cite{thoudam}, for example).  
The possibility  that the  observed anisotropy  can be associated with a structure of  local intersellar magnetic fields
is not discussed in the paper.

\section {The evaluation of a cosmic ray anisotropy  and flux from a local source}
  Experimental data on the energy dependence of an amplitude of  dipole anisotropy in the energy range from 100 up to $ 10 ^ 5 $ GeV are shown in Fig. 1. 
The question to answer is:  what sort and value of anisotropy one should  expect after   a nearby supernova explosion?
  The amplitude of the dipole anisotropy  $ A $ is given by:
$ A = \frac {3D} {2c} \times \frac {1} {n} \left|\frac {dn} {dr} \right| $ \cite{ginzburg},
and for a point-like  instantaneous source it is equal to
      $ A = \frac {3r} {2 \times c \times t} $ where $r$ is the distance from a source to an observer and $t$ is the time of propagation of particle released to the interstallar medium 
at the time $t=0$.
 The anisotropy of a total flux from a local source and the background flux is given by
$ A (E) = A_s \times I_s (E) / (I_s(E) + I_ {bg} (E)) $
where $ I_s (E) $ and $ I_ {bg} (E)  $ are the fluxes from a point-like  instantaneous source and the background isotropic flux from
the rest of supernovae.

We consider that the dependence of the diffusion coefficient on the proton energy is $ D = D_0 \times E ^ {\delta} $, 
where $\delta $ determines the
leakage from the region of confinement of cosmic rays. The value of $\delta$  is known from the  ratio of the secondary and primary nuclei at low energies
(below 100 GeV per nucleon) and is equal to $ \delta = 0.6 $.
(We consider only protons, and for $E>100 $ GeV we replace the rigidity dependence  to the energy dependence.)
The solution for a local instantaneous source  at the distance $ r $ from an observer is given by the Green's function:
$ n(r, E, t) = \frac {1} {(4 \pi \times D (E) \times t) ^ {3/2}} \times exp (-r ^ 2 / (4 \pi \times D (E) \times t)) \times Q (E) $
where $ n (r, E, t) $ is the cosmic ray density,
 and $ Q (E) = Q_0 \times (E/E_0) ^ {- \alpha} $ is the source spectrum.
The flux is given by the formula:
                      $ I (r, E, t) = c / 4 \pi \times n (r, E, t) $

 The energy released by a source into the  cosmic rays is equal to:
$ W_{cr} = \int_ {E_ {min}} ^ {E ^ {max}} Q (E) \times E \times dE $

\section{ The background flux intensity}
We can evaluate the background flux of cosmic rays following our three-component model \cite{ZS} based on the combined data from the ATIC-2 and other experiments, including the ones for the all particle spectrum. In that model we described the spectra of protons, helium and heavier nuclei in the energy range of 10 GeV to $10^8$ GeV by means of  three types of sources with different source spectral indexes and the different maximal rigidity under acceleration. 
The  paper  presents  a new view of the origin of that components.
The recent investigations of gamma-rays from supernova remnants demonstrate that the particle spectra in supernova shells are softer than it was previously considered, with the value of the spectral index being  2.3 \cite{caprioli}.
Taking  this fact into account, we consider  the background particle spectrum as a spectrum of particles which have  been advected downstream, suffered adiabatic losses, and, finally released at the death of the remnant.
So, we proposed that supernovae in the Local Bubble generate particles with the index of the source spectrum being $\alpha=2.3$, the maximal rigidity being higher that 100 TeV, and the leakage from the Local Bubble being described by  the value of $\delta=0.6$ (red dotted line in fig.2).   
We propose that additional background flux is the mean background flux of our Galaxy. It is formed by particles which escape from numerous other bubbles to the Galaxy, which we consider as almost closed or with leakage with very weak energy dependence. We have used the value of $\gamma=2.4$ for them (red solid line in fig.2).
This assumption is based on the ATIC data on the sub-iron to iron nuclei ratio \cite{TiFe}, \cite{ClosedGalaxy}. 
As it was shown in \cite{ZS}, the component with $\gamma=2.4$ and $ E_{max} = 4$ PeV can  describe well the all-particle spectrum up to  the knee and in the knee region.

\begin{figure}[h]
\begin{minipage}{18pc}
\includegraphics[width=18pc]{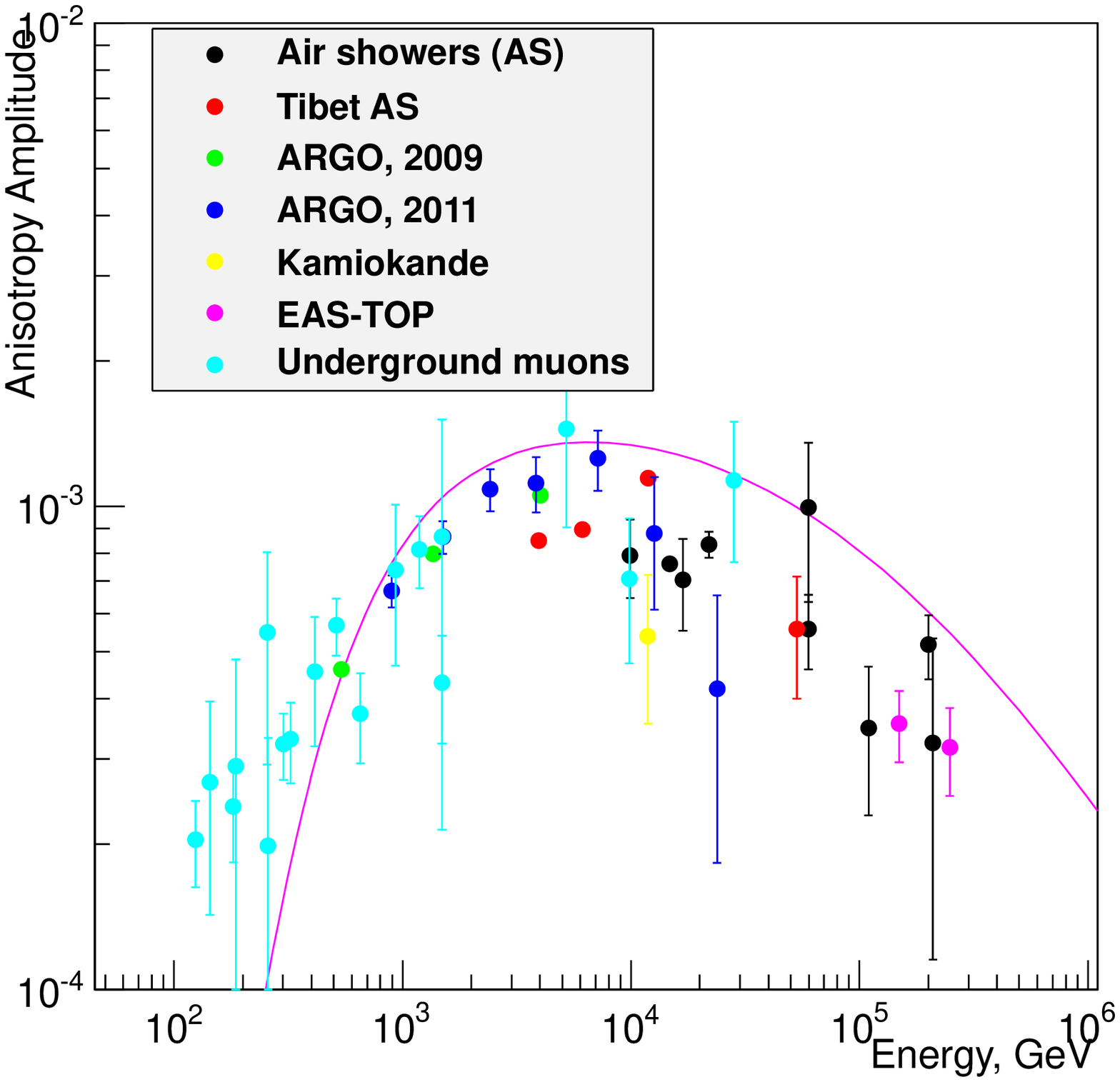}
\caption{\label{fig1} Energy dependence of dipole anisotropy amplitude for $10^2$ -- $10^6$ GeV.
Experimental data: Air showers(AS), Tibet AS, Underground muons: compilation from \cite{tibet}; 
ARGO,2009 \cite{ARGO2009}, ARGO,2011 \cite{ARGO2011}; Kamiokande \cite{kamiokande}, EAS-TOP \cite{KASKADEGRANDE} 
Solid line is the model of a local instantanious source. The parameters of the model are in section Results}
\end{minipage}\hspace{2pc}%
\begin{minipage}{18pc}
\includegraphics[width=18pc]{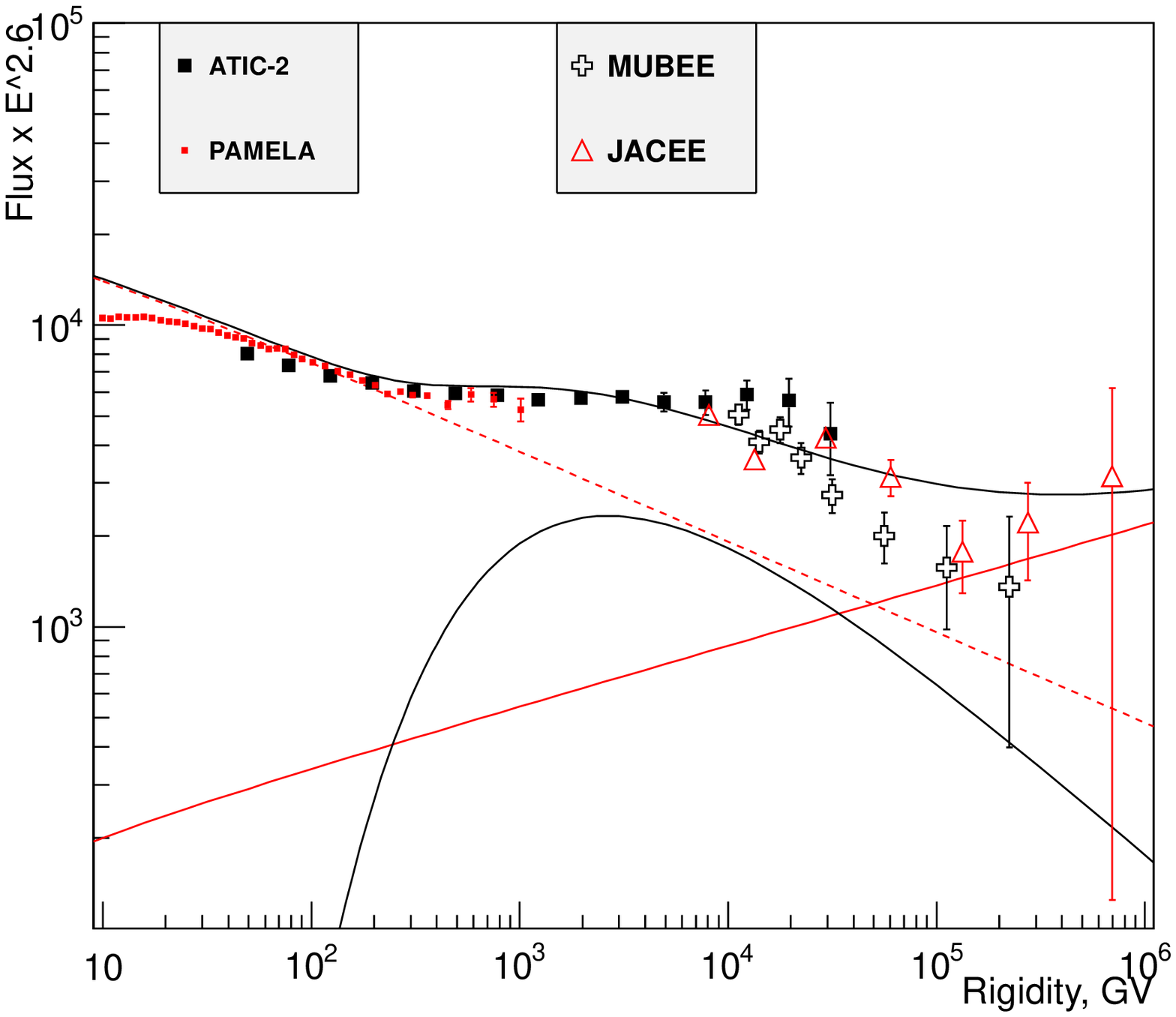}
\caption{\label{fig2} Proton energy spectrum as measured in the direct experiments and its fit to the model of local instantanious source.
Experimental data: PAMELA \cite{pamela}, ATIC-2 \cite{atic}, MUBEE \cite{mubee}, JACEE \cite{jacee}. Red dashed and solid lines are for background spectra, black lines
 -- spectrum of single source and summary spectrum}
\end{minipage} 
\end{figure}

\section{Results}
In Fig. 1 we show  how  to explain the experimental data on cosmic ray anisitropy in the TeV energy range by adding a local instantaneous source.
The model curve is calculated for the following parameters of a local source: 
$ t = 3 \times 10 ^ 4 $ years, 
$ r = 18.4 $ pc,
$ D_0 = 0.6 \times 10 ^ {25} $ cm $ ^ 2 $ s $ ^ {-1} $,
$ Q = 2.5 \times 10 ^ {53} \times (E/1GeV)^{-2.3} $ GeV $ ^ {-1} $, ie $ W_ {cr} \sim 10 ^ {51} $ erg.
For the parameters selection, the important restriction was the one on the source power,
that ought to be not too great to satisfy the experimental estimates.
The proton spectrum for the given set of parameters is shown in Fig.2.
This set of parameters is  of course not unique.   As another example, we have considered the SN wich produced 
the Geminga pulsar. The age of this pulsar is known to be $3.42\times 10^5$ years and its distance is estimated as 280 pc. In paper \cite{gehrels} the Geminga supernova was
 considered as a 
possible cause of the local interstellar bubble. In this case, the power of this supernova explosion should be very high. Because of  asymmetric explosion,
 the pulsar could be 
ejected at a high speed.  In \cite{gehrels} the pulsar motion was taken into account, and it was estimated that the explosion occured about 60 pc from the Solar System.
 The evaluations of the anisotropy and proton spectrum for $t=3.42\times 10^5$ years,
r = 100 pc, and $D_0 = 2.5\times 10^{25}$ cm $ ^ 2 $ s $ ^ {-1} $ give results on anisotropy amplitude and proton spectrum very similar to those shown in Fig.1 and Fig.2, 
providing the source power  is $W_{cr} \sim 2 \times 10^{53}$ erg.  
\section {Discussion}
In  models of particle   propagation the energy dependence of diffusion coefficient is
determined from the ratio of the secondary and primary nuclei and is given by $ D (E) = D_0 \times E ^ {\delta} $.
In an assumption of a stationary picture, where and when supernova explosions in  Galactic disk occur regularly in time and space,
and the particle leakage occurs from the Galactic Halo of the size of several kpc, $ D_0 \sim 10 ^ {28}$  cm$^{2}$  s$ ^ {-1} $ at $ E = $ 1 GeV.
The estimated  diffusion coefficient in our model is of the three orders of magnitude smaller than a normally agreed value  for the Galaxy.
We suggest that the small value of the diffusion coefficient may characterize the smaller value of the leakage region (Local Bubble of $\sim 100 pc$) and 
 the density of the interstellar medium in the Local Bubble that is lower than the mean value in the Galaxy.
 
The discussed model is essentially based on the measured  proton spectrum hardening above 200 Gev and its following  steepening  above 10 TeV.
If future, more precise, measurements  do not confirm this feature, our model of the local source contribution  will have to be rejected.
 
As a consequence of the fast increase in anisotropy with the rigidity, the explanation for cosmic ray  anisotropy above $10^6 GeV$ is possible, 
as due to the 
fluctuations in random
occurence of supernovae in space and time \cite{blasi}. 
But that explanation with given $ \delta = 0.6 $ in formula
$ A = A ^ 0 \times E ^ \delta $
  demands the  value of $ A^0 = 4\times 10 ^ {-7} $. 
This result is shown in Fig. 3 together with the experimental
data and agrees  satisfactorily with the
experiment at the  energies higher than some $10^5$ GeV.
For the confinement region of the size  of 100 pc this corresponds to the value of the diffusion coefficient  $ \sim 10 ^ {25}$ cm$^ 2$  s$^{ -1} $ at 1 GeV,
that agrees well whith the diffusion coefficient we found for the Local Bubble.
It is quite possible that with the  increasing accuracy of available experimental data  this description  will become insufficient and some additional suggestion will be needed,
such as an additional  local source. The additional local source could  explain, as well, the  all-particle spectrum  hardening in the energy region above
$ 2 \times 10 ^ {16} $ eV observed in the recent experiments \cite {kaskade}, \cite {tunka}.

The work was supported by RFBR grant number 11-02-00275.
\begin{figure}[h]
\includegraphics[width=22pc]{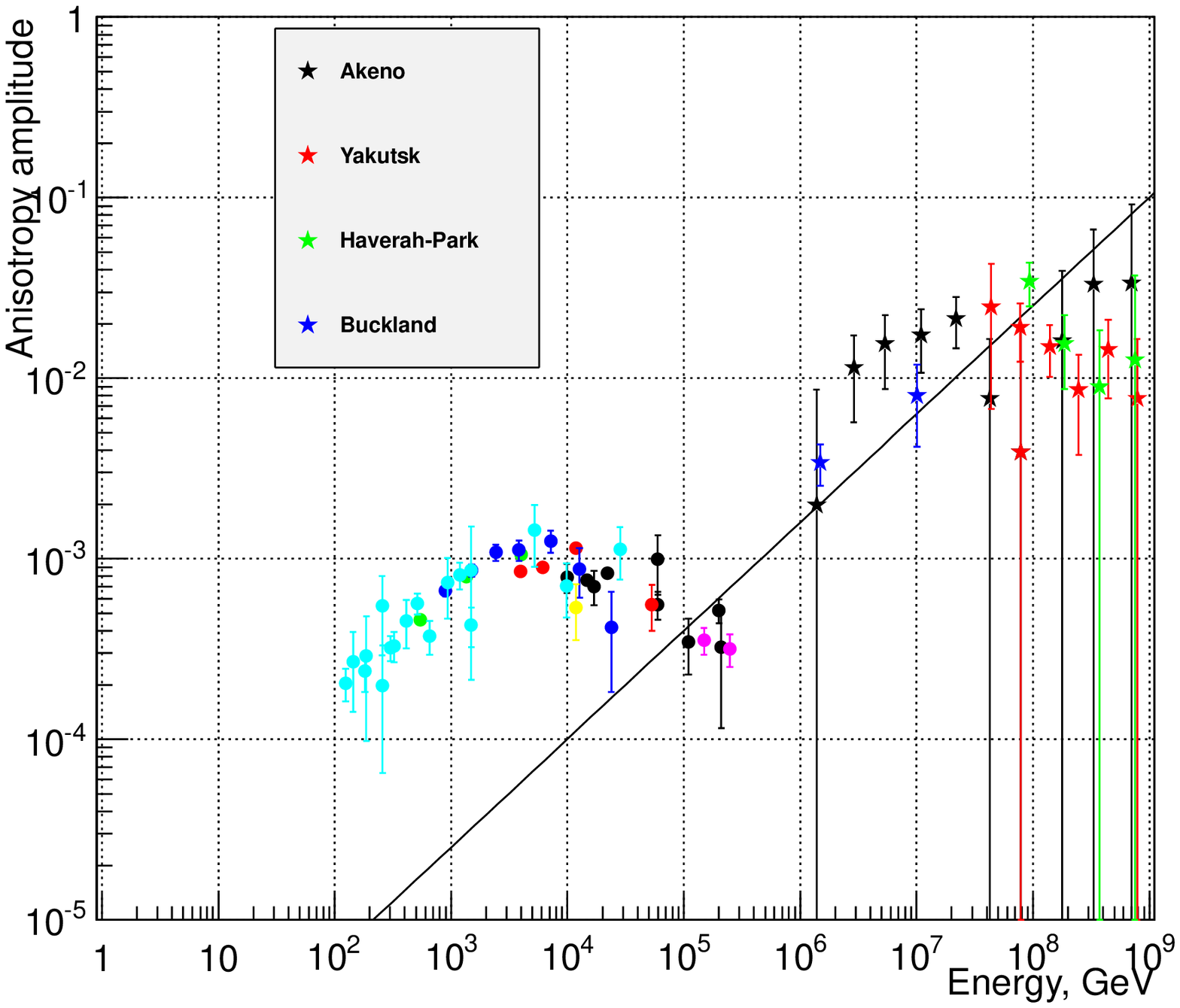}\hspace{2pc}%
\begin{minipage}[b]{12pc}\caption{\label{fig3}Energy dependence of dipole anisotropy amplitude above  $10^6$ GeV.  Experimantal
data of  Akeno, Yakutsk, Haverah Park, Buckland (compilation from \cite{KASKADEGRANDE}) are added to the data of Fig.1. Black line shows increase of anisotropy amplitude from $A = A^0 \times E^{0.6}$ 
with $A^0 = 4\times 10^{-7}$. }
\end{minipage}
\end{figure}
\section *{References}

\smallskip

\end{document}